# The Relationship between Organization Culture and Knowledge Sharing towards Business System Success

**Mohammed. I. Alattas**
Faculty of Engineering and Information Technology
University of Technology Sydney
Sydney, Australia
Email: mohammed.i.alattas@student.uts.edu.au

**Kyeong Kang**
Faculty of Engineering and Information Technology
University of Technology Sydney
Sydney, Australia
Email: kyeong.kang@uts.edu.au

## ABSTRACT

Understanding the business systems success factors has been a challenging process for both public and private organizations. Organizational culture is measured as a critical factor promoting knowledge sharing among employees. Based on the competing value framework (CVF), this research shows how various dimensions of organization culture influence knowledge sharing towards business systems success at the individual and organization level. A quantitative approach was applied to test the relationship between organizational cultures, knowledge sharing and business systems success in Saudi enterprise.

## INTRODUCTION

The importance of the organization culture is critical to the success of business systems usage. There is even more need in Saudi Arabia organizations where a number of other factors can influence the success of business systems success such as knowledge sharing. The academic debate on the role of culture within the business environment highlights the crucial role of culture as a way of improving the new business systems (Juntiwasarakij 2008). Stakeholders need to ensure that the culture of the organization is well understood, particularly in the Saudi context, which is often problematic (Adlan and Have 2012; Al-Adaileh and Al-Atawi 2011; Eid and Nuhu 2011). The cultural setting of Saudis is actually Arab and Muslim. It is widely known that the Saudi setting has a unique culture and heritage, which has been preserved since the inception of the culture (Eid and Nuhu 2011). Visitors to Saudi, including non-Saudis, are subjected to the same rigorous Islamic law as Saudis. In this regard, different enterprises in Saudi are to a greater extent influenced by the cultural aspects of the Saudi community. When it comes to hiring employees in the enterprises, there will be a clear stipulation that employees, whether of Saudi origin or otherwise, will be governed by similar policies and will follow similar requirements for their enterprises. Differences in culture exist in Saudi enterprises, and theses differences in values, beliefs and customs will affect the diverse employees enterprises in a working atmosphere (Aleisa and Diboqlu 2010).

Understanding the success factors for business systems has been a challenging process for both public and private organizations in Saudi Arabia. Business system success is complex in nature as it connects each functional departments of an entire firm, may take several years to go live and requires committing a significant amount of a budget and other resources (Raymond et al. 2005). In order to increase the business system success rate, organization culture has been recognized as one of the most important success factor in the literature (Gou 2012). The impact of organization's culture on organizational actors can also mean the actors are able to not only implement the systems, but also collaborate to improve their efficiency and effectiveness. Collaboration within the organization is also considered to represent a crucial aspect affecting the overall performance of a company (Boehm 2012). Collaboration is one of the most important factors for the success of modern organizations, which can ensure they have the most robust mechanism available to them (Alston and Tippett 2004; Beauregard 2011). Collaboration needs to ensure by organizational actors in order to lead a competitive advantage for the firm (Crow and Hartman 2002). Therefore, it is important for organizational managers to have a clearer understanding of the needs of the organization and the success of business system success, which is increasingly important for the organization (Liu et al. 2007). The new business systems of Saudi organizations need to improve considerably and therefore there is a need for organizations to ensure collaboration can be enabled (Alston and Tippett 2004, 2009; Beauregard 2011). Much research has been done in the area of assessing information system success (Popovič et al. 2012), with the DeLone and McLean (1992) multidimensional Information Success Model for organizing the concept of information system success as being one of the most often used works.

Organizations need to overcome the cultural barriers and initiate appropriate culture to best facilitate knowledge sharing (Jones 2005; Jones et al. 2006). Business systems such as ERP usually comprise of integrated modules across multiple business functions, therefore distribution of organizational knowledge is significant in building intensive knowledge platform and providing cost-effective functionalities"(Hendricks 2007). Although the existing literature has examined the "link between organizational culture and knowledge sharing (Jones 2005; Jones et al. 2006), and their relationship with ERP success (McGinnis and Huang 2007), little research has focused on understanding the influence of organizational culture and knowledge sharing towards business system success in Saudi enterprises.

Therefore, it is critical to identify various factors of business systems for a successful outcome in a Saudi firm. To achieve the goal of this research, the following research questions are addressed: (1) how does four cultural types (group culture, hierarchical culture, development culture and rational culture) impact business system success in Saudi enterprise? (2) How does knowledge sharing influence business system success in Saudi enterprise? The study is organized as follows. The following section provides a literature review. Then in the next section, a theoretical development and research model are presented. Followed by data analysis. The last section concludes the study and present future research.

# LITERATURE REVIEW

## Business System Success

Business systems are examples of large complex Information Systems (IS), which are integrated throughout cross-functional departmental boundaries in an organisation (Brady et al., 2001). Business systems have been defined as business software systems that let an organisation (Umble et al., 2003):
- To assimilate and automate and organisation business processes.
- To share everyday data and information throughout the organization.
- To access and generate information in a real-time situation.

Prior studies investigated how to evaluate an IS (such as business systems) from a business perspective. For-example, Shang and Seddon (2002) highlighted that various stakeholders and end-users have diverse system perceptions and needs. Success is referenced against various criteria, such as organization goals, financial performance and on-time delivery (Markus and Tanis, 2000). Numerous models have suggested that explain dimensions of business system success. Though, the most commonly cited model is the IS Success model by DeLone and McLean (1992). The model highlights the "understanding of the connections between the different dimensions of information systems success. Such as six major dimensions of IS success – System Quality, Information Quality, Use, User Satisfaction, Individual Impact, and Organizational Impact".

Sedera et al. (2004) denoted that the most significant enterprise system success dimensions are Organizational and Individual Impact. Business systems are measured successful at the post-implementation phase, if it improves potential benefits through firm cost reductions, increased customer satisfaction levels, higher operational productivity etc. (Sedera et al., 2004). It is worth mentioning that other investigators have used the Sedera et al. (2004) model of ERP success in their studies (such as Ifinedo, 2007; Sehgal & Stewart, 2004; Wang et al., 2008; Yoon, 2009). The focus of business systems research so far has been on the adoption and implementation phase (Umble et al. 2003). Most of extant studies assess ERP success by whether the system is implemented on time and within budget, but ignore that the ultimate goal of using business systems is to create business value and enhance business performance (Shao et al. 2012).

## Organization Culture

Organizational culture is considered as a critical factor promoting collaboration between staff, in particular knowledge sharing (Shao et al. 2012). According to (Eid and Nuhu 2009) organisation culture effects employee collaboration, organizational functioning, and even decision making in organizational settings. Organizational culture is the encouraging factor that makes a business profitable by leading employees to acquire knowledge and develop innovative ideas (Hahn et al., 2013). Mueller (2014) investigated the cultural context of knowledge sharing between project teams and found that learning culture supports knowledge processes and employees see knowledge sharing as a natural activity in their daily business. Škerlavaj et al. (2010) describes organizational culture as a complex process that refers to the development of new knowledge and has the potential to change individual and organizational behaviour. According to (Škerlavaj et al. 2010) within the competing values framework (CVF) organization learning culture has four different types of cultures: group, developmental, hierarchical, and rational. Figure 1 show CVF model.

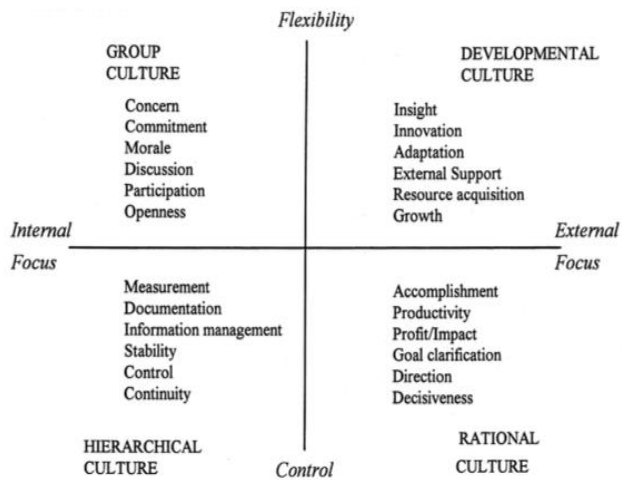

Figure 1: Denison and Spreitzer, (1991) organizational culture

The competing values framework (CVF) explores the competing demands within an organization on two axes (Denison and Spreitzer 1991; Quinn and Spreitzer 1991; McDermott and Stock 1999). The first dimension, the flexibility–stability axis, reflects the competing demands of change and stability. The second dimension, the internal–external axis, focuses on activities happening within or outside the organization. The two axes divide organizational culture into four culture domains: a group culture, a developmental culture, a rational culture and a hierarchical culture. The group culture emphasizes flexibility and maintains a primary focus on the internal organization. Belonging, trust, attachment, cohesiveness, and participation are core values. The development culture also emphasizes flexibility and change, but maintains a primary focus on the external environment. Growth, resource acquisition, creativity, stimulation and adaptation to the external environment are core values. The rational culture emphasizes internal stability and external environment. Planning, efficiency, productivity, goal fulfilment, and achievement are core values. The hierarchical culture focuses on internal organization and stability. Internal efficiency, coordination, order, rules, control and regulations are core values.

## Knowledge Sharing

In the business environment, information remains a vital element in ensuring optimal performance of the different entities within an organization. This involves collaboration between departments, employees, management and all the internal staff of the company. Effective internal collaboration within companies ensures employees remain focused and engaged in delivering various organizational goals. Collaboration both within the organization is often considered to represent a crucial aspect affecting the overall performance of a company (Boehm 2012). In essence, the practice of sharing information, experiences and resources is the key to future development, and information systems have been shown to play a vital role in enhancing the level of collaboration (Alston and Tippett 2009; Boehm 2012). Although this may be an understood concept in private organizations, in many organizations in Saudi these factors have not been well understood, and therefore can create a number of issues for organizational actors. Teamwork is also critical to the different organizations. They continue to work together to improve their effectiveness and collaboration, as part of the culture of the organization enables improvements in the efficiency of the organization (Shao et al. 2012).

Knowledge is the foundation of a firm's competitive advantage, and, ultimately, the primary driver of a firm's value (Kraaijenbrink 2010). Such as Knowledge sharing: explicit knowledge sharing and tacit knowledge sharing. According to (Shao et al. 2012), "explicit knowledge is formal and systematic, and can be achieved through readings of project manuals and team discussions, while tacit knowledge is highly personal, context-specific, subjective, and can be represented in the form of metaphors, drawings, non-verbal communications and practical expertise. It is usually difficult to articulate tacit knowledge through a formal use of language since it is expressed in the form of human actions such as evaluations, attitudes, points of view, commitments and motivation".

## THEORETICAL DEVELOPMENT AND RESEARCH MODEL

Business systems are considered information system (IS), which are integrated throughout cross-functional departmental boundaries in an organisation (Umble et al., 2003). Therefore, prior research on user acceptance models for information systems is useful to recognize the success of business systems success. This study deals with widespread models related to information system (IS) acceptance, which are the DeLone & McLean (D&M) IS Success Model, the Business Systems Success Measurement Model (Gable et al., 2003; Sedera et al. 2004). In addition, Organization culture based on Competing Values Framework (CVF) (Quinn and Spreitzer 1991; Denison and Spreitzer, 1991; and McDermott and Stock, 1999) is reviewed for identifying the factors affecting business systems success.

Saudi Arabia has managed to attract a workforce from all over the world, and this has enabled it to merge varied cultures. The business system success at the individual and organization level can be improved if the blow factors to be taken into account. From the above review and culture competing values framework (CVF) model, using our research model we seek to show the impact of organizational culture and employee's knowledge sharing towards business systems success. Figure 2 shows our research model. Table 1 shows the definition of each factor used in the current study.

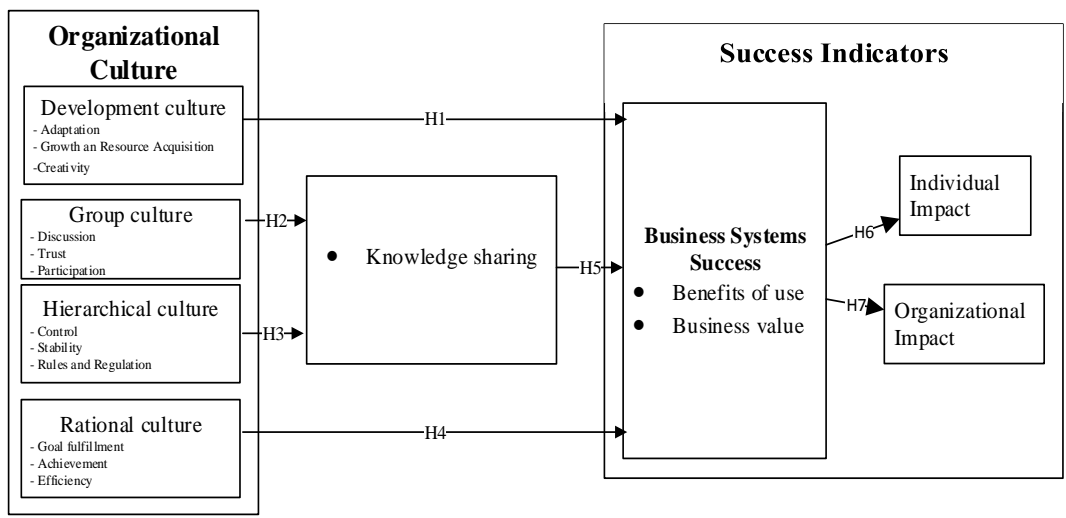

Figure 2: Research model.

| Factors | Sub dimensions | Description | Sources |
|---|---|---|---|
| **Development Culture** | Innovation | The flexibility of organization towards change and encourages innovation. | (Guo 2004; Shao 2012; McDermott and Stock 1999; Škerlavaj et al., 2010) |
| | Adaptation | The top executive needs to facilitate a development culture that focuses on innovativeness, creativity, and adaptation to the external environment, such as the organization "would tend to scan the competitive environment to assess their relative competitive strengths and weaknesses in relation to their competition and customers, and strive to make changes to their firm accordingly". | |
| | Growth and Resource Acquisition | | |
| | Creativity | | |
| **Group Culture** | Discussion | Employees share information and insights throughout the organization and have considerable influence over decision-making. | |
| | Trust | The top executive needs to promote a trust-oriented group culture that focuses on | |
| **Hierarchical** | Participation | belonging and participation. | |
| | Control | Firm standardization to achieve control | |
| | Stability | Internal firm desire for a focus on change or stability; firm emphasizes on stability. | |
| | Rules and Regulation | A concern with formal rules and procedures. | |
| **Rational Culture** | Efficiency | The degree of importance placed on employee efficiency and productivity at work. | |
| | Goal fulfilment | A concern with clearly defining the goals of the organization. | |
| | Achievement | Firms emphasize on productivity and achievement, with objectives typically well-defined and external competition a primary motivating factor. | |
| **Knowledge sharing** | **Information sharing** refers to the extent to which a firm shares a variety of relevant, accurate, complete, and confidential information in a timely manner. "Sharing of knowledge about business processes and the related knowledge required to make these processes work" | | (Jones et al. 2006) |
| **Business system Success** | **Benefits of use and Business value:** establishes the extent to which business systems are contributing to the success of the different stakeholders. Net benefits: as they capture the balance of positive and negative impacts of the business system on organizations. | | (Chien and Tsaur, 2007) |
| **Individual Impact** | The measure of the effect of information on the behaviour of the recipient. | | (DeLone and McLean 1992; Ifinedo 2010) |
| **Organizational Impact** | The measure of the effect of information on organizational performance. | | |

Table 1: Description of terms used in our research model

## Hypotheses Development

**Organization Culture**

**Development Culture:** The role of development culture within the business environment highlights the crucial role of culture as a way to business systems success. The employees' innovativeness has an effect on the success of business systems, as culture is one of the defining characteristics for any organisation today (Rashid et al., 2003). Organisational Culture has been indicated to effect technology change within organisations (Hannan and Carroll, 2003). In order to improve and achieve business systems success, "the top executive needs to facilitate a development culture that focuses on innovativeness, creativity, and adaptation to the external environment, thus to offer the users a vision of organizational strategic directions and inspire the users to think innovatively about how the system might enable the business to accomplish its goals and achieve business performance" (Shao et al. 2012). This leads to the following hypothesis.

Hypothesis (H1): Development culture has a significant positive impact on business systems success in Saudi enterprise.

**Group Culture:** Liu et al., (2010a) reported that continuous learning by staff is one of the important activities in enterprise system adaptation. Since business systems assimilate several business functions, employees must not only be aware of their own job and responsibility, but also collaborate thoroughly with other employees in organizational primary business process (Liu et al., 2010b). Ke and Wei, (2008) reported that the employees interaction with organizational members support knowledge gathering and thus this culture of knowledge sharing helps to increase employees' confidence and reduce their fear to share their knowledge. In order to motivate employees' to learn systems functionalities and facilitate organizational sharing of business system knowledge, the top executive needs to promote a group culture that focuses on participation by taking account of their individual needs (Shao et al. 2012). Jones et al. (2006) discovered that organizational culture that emphasizes on teamwork, and collaboration can facilitate knowledge sharing in enterprise systems. In particular, group culture enable tacit knowledge sharing within the organization (Jones, 2005). This leads to the following hypothesis.

Hypothesis (H2): Group culture has a significant positive impact on knowledge sharing in Saudi enterprise.

**Hierarchy Culture:** Lin (2007) suggested that certain forms of extrinsic motivation such as incentives or praise and public recognition might stimulate staff individual motivation and foster their knowledge sharing intention. In order to promote individuals' active participation in business systems training, the top executives need to set up suitable evaluation mechanisms and organize a system of reward mechanisms to raise a hierarchical culture that emphasizes efficiency and coordination (Podsakoff, et el., 2006; Sharma and Yetton, 2003; Umble et al., 2003). Thus lead to

Hypothesis (H3): Hierarchy culture has a significant positive impact on knowledge sharing in Saudi enterprise.

**Rational Culture:** Jones et al. (2006) examined that organizational rational culture is positively related with business systems with organisation. An organisation implementing with a high degree of external orientation is more likely to achieve business success (McDermott and Stock 1999). For-example, organisations that emphasizes an external orientation (rational culture) is more likely to experience positive competitive outcomes. Organisations need to ensure they have the best cultural support available for the success of the business systems (Kaptein 2011; Zhang 2010). Organisation needs to promote a rational culture (Shao et al. 2012). This leads to the following hypotheses:

Hypothesis (H4): Rational culture has a significant positive impact on business systems success in Saudi enterprise.

**Knowledge Sharing**

Knowledge-sharing within the organization is considered a vital factor impacting the overall performance of an organization (Boehm 2012). In essence, information systems have been shown to play a vital role in enhancing the level of knowledge sharing (Alston and Tippett 2009; Baird 2012; Boehm 2012). Gable, Scott, and Davenport (1998) suggested effective knowledge sharing in particular offer significant commercial and practical benefits to a business system success. Knowledge transfer maintains the organization and evolves its business system to generate returns (Davenport 2000). Bock et al. (2005) suggested that employees are more likely to share knowledge with their coworkers in trust-oriented culture to form a mutual belief that focuses knowledge attainment inside the organization, which are significant factor of business system success (Vandaie, 2008). Employees need to experience a continuous learning process to build a strong relationship between what employees have known and what the business systems wants them to know (Ravichandran, 2005; Ke and Wei, 2008).

Knowledge sharing is considerably important for organisations to ensure they have the best possible system available to them, which can ensure the long-term success (Kratzer et al. 2011; Roggeveen et al. 2012). Wang et al. (2007) showed that the active knowledge-sharing could produce a better relationship between business systems and organizational processes to improve business performance for achieving competitive success. Thus we hypothesize.

Hypothesis (H5): Knowledge sharing has a significant positive impact on business systems success in Saudi enterprise.

**Businesses System Success**

Business systems provide a backbone of information, interaction and control for an organization (Shehab et al., 2004). Zhu et al. (2010) highlighted that the business systems directly impact the managerial and operational processes. Therefore benefits resulting from improvement in those managerial and operational processes can improved the direct benefits to the organization (Shao et al. 2012). Liu et al. (2011) discussed that individuals' and their ability to use enterprise systems and their understanding is critical for organizational level ERP adaptation. From organisation perspective, a successful business system reduces uncertainty of results and thus lowers risks, and controls inadequate resources (Chien and Tsaur 2007). From the end user's perspective, a successful business system is to improve the user job performance without frustration. Individual impact refers to measuring the influences brought by the business system on individual users, such as changes in productivity and decision-making. Kositanurit et al. (2006) also found a significant positive relationship between ERP system and individuals performance of using such systems. Organizational impact requires the evaluation of the changes caused by the business system to the organization, such as increase or decreases in operating costs and growth in profits etc. (Chien and Tsaur 2007). Thus the following hypotheses are developed.

Hypothesis (H6): Business systems success increases organisation impact in Saudi enterprise.

Hypothesis (H7): Business systems success increases individual impact in Saudi enterprise.

## METHODOLOGY

This research applied quantitative method to collect numerical data from participants in Saudi firm using business systems such as Oracle e-business applications. The survey instrument is used for this study to collect data. Data collection lasted from November 2014- March 2015. This study adopted previously validated instruments in order to ensure the measures are adequate and representative. Appendix A shows all item used in the study. The scales implemented in this survey were developed originally in English. However, a certified translator translated the English version to Arabic. The five point Likert scale (1=strongly disagree to 5=strongly agree) are used because it is considered one of the most commonly used techniques of scaling responses in a survey design. Data was collected in 'Saudi Binladin Group', which is one of the Saudi biggest enterprises. Total of 500 participants were contacted, 350 participated in the survey. After removing incomplete responses, in total 330 responses were collected. The Partial Least Squares (PLS) technique (using SmartPLS version 2.0) was used to test the research model. Partial Least Squares such as structural equation modelling (PLS-SEM) is used to estimate the relationships between the different parameters of the research model. Partial Least Squares (PLS-SEM) structural equation modeling (SEM) tests theoretical models to understand the simultaneous modeling of relationships among various independent and dependent variables.

## RESULTS

Descriptive analysis shows that that majority of the participants were male 209 (63.3%) and 121 (37.7%) are female. 46.9% were of 26-35 years old, followed by 33.3% participants were 36-45 years old. The majority of participants hold the bachelor's degree with 51.5%, followed by Masters degree with 35.7%. 50% of participants had 3-5 years of work experience, followed by more than 3-5 years (29%). Then 18% of participants had work experience of 1-3 years. 50% are in supervisory level of management followed by 36.7% was in mid-level of management and 13.3% were in top level of management. (29.6%) of participants were from human resource department, 21.8% participants were from finance and accounting, followed by 24.2% from sales and marketing. 18.1% were from production and procurement department. In addition, 6.3% belong to other departments. In the HR department, the high percentage 61.2% used 'oracle human resource'. In finance and accounting department, 62.5% used 'Oracle Payroll (PAY) or Account Payables'. In sales and marketing department, 28.7% used 'Oracle sales' while in production/procurement department 41% used 'oracle order management'.

### Reliability and Validity Assessment

In our research model, all factors were modelled as reflective indicators because they were observed as effects of latent variables. Internal consistency, convergent validity and discriminant validity were evaluated. To demonstrate the reliability of the latent variables, the loadings of individual measures for each variable exceeded 0.7 values and were significant at p value < 0.05. Cronbach's alpha (CA) and the composite reliability (CR) coefficients exceeded an acceptable measure 0.7. Convergent validity is assessed using average variance extracted (AVE) and the CR. Convergent validity is established if the CR value is more than the AVE and all the AVE are greater than 0.50 (Hair et al., 2006). Discriminant validity assesses whether a construct is different from all other constructs; the square root of individual AVE should be more than any correlation between the latent variables (Zait and Bertea, 2011). The square root of the AVE for each individual constructs is greater than the correlations with all other constructs. This demonstrates the discriminant validity of all the constructs in the study. Table 2 shows the reliability and validity results.

|  | CAlpha | AVE | CR | DC | GC | HC | RC | KSS | BSS | OM | IM |
|---|---|---|---|---|---|---|---|---|---|---|---|
| DC | 0.81 | 0.81 | 0.71 | **0.90** | | | | | | | |
| GC | 0.80 | 0.79 | 0.85 | 0.33 | **0.89** | | | | | | |
| HC | 0.81 | 0.80 | 0.81 | 0.77 | 0.65 | **0.89** | | | | | |
| RC | 0.82 | 0.81 | 0.86 | 0.76 | 0.58 | 0.63 | **0.90** | | | | |
| KS | 0.85 | 0.80 | 0.76 | 0.06 | 0.08 | 0.45 | -0.06 | **0.89** | | | |
| BSS | 0.87 | 0.77 | 0.79 | 0.17 | 0.19 | 0.05 | -0.17 | 0.65 | **0.87** | | |
| OM | 0.82 | 0.81 | 0.77 | 0.05 | 0.14 | 0.20 | -0.05 | 0.67 | 0.55 | **0.90** | |
| IM | 0.84 | 0.79 | 0.76 | 0.71 | 0.65 | 0.77 | 0.71 | 0.62 | 0.58 | 0.66 | **0.89** |

- AVE: Average variance extracted, CR: Composite reliability, DC: Development Culture, GC: Group Culture, HC: Hierarchy Culture; KS, Knowledge sharing; BSS: Business System Success, KS, Knowledge sharing; BSS: Business System Success, OM: Organisation Impact, IM: Individual Impact.

Table 2: reliability and validity.

## Structural Model Testing

After following the acceptable measurement model as discussed in above sections, the structural model testing was done to test the hypotheses proposed in the current study. The following sections explain the structural model for latent constructs to answer the research questions and address the hypotheses. Table 3 show the path co-efficient mean, standard deviation and t-statistics and p-value for each of the proposed hypotheses. The recommended value are t >1.96 at p < 0.05, t > 2.576 at p < 0.01, t > 3.29 at p < 0.001 for two-tailed tests. Figure 3 shows the path testing.

| | Path | Path coefficent | StDev | T statistics | P values | Supported? |
|---|---|---|---|---|---|---|
| H1 | DC -> BSS | 0.25 | 0.03 | 1.98 | 0.003* | Yes |
| H2 | GC -> KS | 0.43 | 0.05 | 4.79 | 0.000*** | Yes |
| H3 | HC -> KS | 0.34 | 0.03 | 3.61 | 0.000*** | Yes |
| H4 | RC -> BSS | 0.145 | 0.03 | 1.20 | 0.336 | No |
| H5 | KS-> BSS | 0.75 | 0.04 | 5.90 | 0.000*** | Yes |
| H6 | BSS -> OM | 0.52 | 0.07 | 2.50 | 0.000*** | Yes |
| H7 | BSS -> IM | 0.24 | 0.03 | 1.99 | 0.002* | Yes |

Notes:
- StDev: Standard deviation, DC: Development Culture, GC: Group Culture, HC: Hierarchy Culture; KS, Knowledge sharing; BSS: Business System Success, KS, Knowledge sharing; BSS: Business System Success, OM: Organisation Impact, IM: Individual Impact
- *Significant at 0.05 level **, Significant at 0.01 level, *** Significant at 0.001 level

Table 3: Hypothesis testing

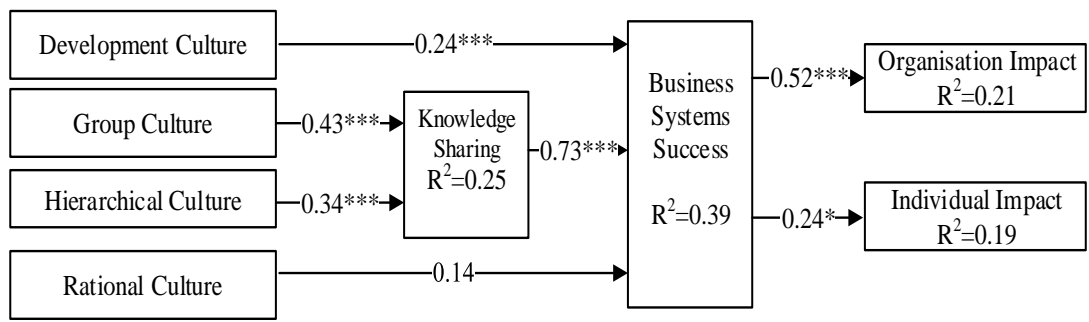

Figure 3: path testing.

## FINDINGS and DISCUSSION

According to the path testing as shown in Figure 3, the order of significance among the organisation culture factors that have a significant effect is "Group Culture", followed by "Hierarchy Culture", "Development Culture" and "Rational Culture". This mean 'group culture' is the most important factor to be associated with knowledge sharing towards business systems success in Saudi context. It can be interpreted that staff share information and insights throughout the organization and have considerable influence over decision-making. The organisation promotes a trust-oriented group culture that focuses on belonging and participation. The reason behind the significance of "Group Culture" is that sharing more accurate data and timely information with others is one of the biggest advantages of business systems success, so users may think that they should use business systems because of their work group. Bock et al. (2005) suggested that because of the trust-oriented culture, employees are more likely to share knowledge with their co-workers, thus to form a shared belief that emphasizes knowledge acquisition and application within the organization.

Another noteworthy finding is that 'hierarchy culture' has a significant impact on knowledge sharing. This indicates that the hierarchical culture focuses on internal organization and stability to emphasize on efficiency and coordination. The reason behind the significance of hierarchy culture' is that sharing knowledge with others is one of the biggest advantages of business systems success, so users may think that internal efficiency, coordination, rules, control and regulations are core organisation values (McDermott & Stock, 1999).

The results show that "Development Culture" has a marginal impact on business system success. This indicates that in order to facilitate a development culture, which focuses on innovativeness, creativity, and adaptation to the external environment, the users should think innovatively about how the business systems accomplish its business performance (Shao et al., 2012). As hypothesized "Rational Culture" has significant positive impact on business systems success in Saudi enterprise. However, its impact is not significant even though it has a strong correlation with other factors. This indicates that if the objectives, productivity of the firm are not achieved through business systems then the users should use it. Rational culture is the development of new knowledge and has the potential to change individual and organizational behavior (Škerlavaj et al., 2010).

Another interesting finding is that knowledge sharing has positive impact on business system success. This indicates that effective knowledge sharing can lead to a business systems success to enhance business performance and achieve competitive success. Knowledge sharing is important for employees to integrate knowledge, thus to have a deeper understanding of business system functionalities and capabilities (Wang et al., 2007). The whole empirical model predicted 25% of the variance in knowledge sharing, 39% in business system success, 19% s individual impact and 21% is organisational impact. Finally, it was anticipated that the success indicators of business system have a positive impact on the final dependent variables "organizational impact" and "individual impact". The results indicate that the more use of business systems can increase the benefits of organization and also the individual impact.

## CONCLUSION and IMPLICATIONS

This study brings new understandings regarding the success of business systems through the inclusion of organization culture and knowledge sharing towards success indicators of business systems in Saudi enterprise. Hence, the outcomes of the study will be of great significance to Saudi enterprise in terms of business systems success. Thus, the result of this study may be relevant to both public and private organizations in the Saudi Kingdom.

### Theoretical Implications

There are several theoretical implications resulting from this study. Firstly, this study has made a significant contribution to the body of knowledge in that it is the first to explore the Organisational Culture – Knowledge Sharing – Business System Success relationship in Saudi context. This study has confirmed that such a relationship exists and has identified key elements of that relationship. Secondly, the main theoretical contribution of the study is the development of a business system success model that can be used in further studies. Therefore, this research contributes to the existing knowledge by proposing business system success model that includes the role of organization culture based on competing values framework (CVF). Thirdly, this study extended prior research on the effects of knowledge sharing on business system success at the individual and organization level. Additionally, this research addresses the shortcomings in the existing literature, by applying knowledge sharing in Saudi enterprise towards business system success. Fourth, this study confirmed that organization culture and knowledge sharing are positive related to business system success. It also shows the positive relationship between organization culture and knowledge sharing in Saudi context. Finally, the various hypotheses supported in this study all add to the literature for developing hypotheses for future studies. Additionally, this study contributes to validating the survey instrument of the various factors used in a proposed model.

### Practical Implications

This study has several practical implications. Firstly, from the managerial perspective, this study provides insights for the Saudi organizations to pay attention to the influence of organizational culture on business system use. Secondly, the managers should not pay attention on only one cultural type, but should focus on all four cultural types (development, group, hierarchy and rational culture) to form a well-balanced culture to achieve success of business systems in terms of organization and individual impact. Thirdly, the top executives should realize that knowledge sharing is important towards the business system success. Therefore, top executives should set up definite rules and regulations, hierarchical structure, and formal communication channels so as to promote the success indicators of business systems. Fourth, the top management who control resources and decision-making must play a significant role in building organization-wide awareness of knowledge sharing practices and how they can contribute to systems improvements. This emphasizes the importance of managers increasing the usage of business systems to increase employee confidence levels and consequently improve organizational performance. Fifth, the top management support and internal incentives should be promoted for effective knowledge transfer within organizations. The interconnection between knowledge management and organizational culture will improve their competitive advantage and increase their business systems' performance. Which in turn improve organization's operations and financial performance. Sixth, organizations are not generally using their business systems as a platform to access information and knowledge. Therefore, knowledge management practices should be incorporated into evaluations of business systems success to help mitigate potential dissatisfaction with business systems investments. Seventh, for both small and large organisations that are planning to apply new business systems will be better able to identify those factors (organization culture and knowledge sharing) that will enhance the possibility of success. The findings of this study will help them to establish those factors on which they should give specific attention to ensure that they receive continuous management scrutiny. Finally, the top executives should set up clear goals and inspires staffs to achieve goals by rational effectiveness criteria, so as to increase employees' perception that organizational practices are equitable and to foster an effective sharing of knowledge in the long-term business system success.

### Limitations and Future work

Like most research, this study has some limitations. First, this study focuses on a limited number of factors for business systems success. More relevant factors such as system quality, service and information quality may be added to improve the understanding of business systems success in Saudi context. Second limitation of this study is the sampling process. The data were collected from two organisations in one city in Jeddah Saudi Arab, which may affect the generalizability of the findings of this study. In addition, having a larger base of survey respondents and interviewees would provide better insight on the issues on a larger scale.

The third limitation is the determinants of organizational culture have not been taken into consideration, which might weaken the conclusive strength of the findings. Future studies could include the factors that influence organizational culture. Moreover, future studies could be conducted in Saudi Arabia that incorporate measurement of national culture into the research model. As knowledge sharing was strongly related to business system success in Saudi organization. Future studies could explore what specific type of knowledge sharing is more effective, tacit, implicit or explicit knowledge sharing? Additional research could attempt to document the knowledge sharing tools through which employees can share knowledge in an organization.

Limitation of the study includes that there is a chance that other important factors exist that that can lead to success or failure of business systems and the fact that these may differ case by case. It is not easy to consider all the possible factors associated with business systems success. This can negatively impact the proposed model. This research should extent further to identify all possible factors related with business systems success.

| **Appendix A: Survey items** | |
|---|---|
| **Development Culture** | (Guo et al 2014; Shao et al. 2012; McDermott and Stock 1999; Škerlavaj et al., 2010) |
| **DC1:** Our firm emphasize on collaboration for business success. | |
| **DC2:** Our firm emphasizes on growth and acquiring new resources for business systems. | |
| **DC3:** Our firm encourages creativity and/or the development of new ideas. | |
| **Group Culture** | |
| **GC1:** Our firm emphasize on group work. | |
| **GC2:** In my firm, people spend time building trust with each other. | |
| **GC3:** In my firm, teams revise their thinking as a result of group discussion or information collected. | |
| **Hierarchy Culture** | |
| **HC1:** Our firm is a very organized place. | |
| **HC2:** Our firm emphasizes on stability. | |
| **HC3:** Our firm emphasizes on rules and regulations. | |
| **Rational Culture** | |
| **RC1:** Our firm emphasizes on tasks and goal accomplishment. | |
| **RC2:** Our firm emphasizes on competitive actions and achievement. | |
| **RC3:** Our firm is a very efficacy oriented place. | |
| **Knowledge sharing** | |
| **KS1:** I am pleased to share (communicate) my work reports on business systems use. | (Jones et al. 2006, Shao et al. 2012;) |
| **KS2:** I would be pleased to communicate business related official documents with other members. | |
| **KS3:** I would like to provide my expertise on business system use. | |
| **KS4:** Collaboration is a key to knowledge sharing on business system use. | |
| **KS5:** Teams must share knowledge in order to take decisions for business system use. | |
| **Business System Success** | |
| **BSS1:** Our firm establishes good relationships with the user community for business systems success. | (Chien and Tsaur, 2007) (DeLone and McLean 1992; Ifinedo 2010) |
| **BSS2:** Our business system satisfies end-user requirements. | |
| **BSS3:** Our firm establishes and maintains a good image and reputation with end-users. | |
| **BSS4:** Our business system enables the organization to respond more quickly to change. | |
| **BSS5:** Our firm ensures that business system projects provide efficiency. | |
| **BSS6:** The results of business system are achieved through focus on the process of gathering knowledge from business system use. | |
| **Individual Impact** | |
| **IM1:** Our business system use enhances individual creativity. | |
| **IM2:** Our business system use enhances higher quality of decision-making. | |
| **IM3:** Our business system use saves time for individual tasks/duties. | |
| **Factor (Organisational Impact)** | |
| **OI1:** Our business system reduces organizational costs. | |
| **OI2:** Our business system improves overall productivity. | |